\documentclass[sigconf]{acmart}
\AtBeginDocument{%
  \providecommand\BibTeX{{%
    \normalfont B\kern-0.5em{\scshape i\kern-0.25em b}\kern-0.8em\TeX}}}

\usepackage{subcaption, listings}
\usepackage{color, colortbl}
\usepackage{stfloats}
\usepackage{tabularx} %
\definecolor{lightgray}{gray}{0.9}
\usepackage{ifpdf}
\usepackage{layouts}

\copyrightyear{2024}
\acmYear{2024}
\setcopyright{acmlicensed}\acmConference[CHI '24]{Proceedings of the CHI Conference on Human Factors in Computing Systems}{May 11--16, 2024}{Honolulu, HI, USA}
\acmBooktitle{Proceedings of the CHI Conference on Human Factors in Computing Systems (CHI '24), May 11--16, 2024, Honolulu, HI, USA}
\acmDOI{10.1145/3613904.3642090}
\acmISBN{979-8-4007-0330-0/24/05}

\begin{document}

\title{AdapTics: A Toolkit for Creative Design and Integration of Real-Time Adaptive Mid-Air Ultrasound Tactons}

\author{Kevin John}
\orcid{0009-0009-5075-4322}
\affiliation{
    \institution{Arizona State University}
    \city{Tempe}
    \state{AZ}
    \country{USA}
}
\email{kevin.john@asu.edu}

\author{Yinan Li}
\orcid{0009-0000-2131-4234}
\affiliation{
    \institution{Arizona State University}
    \city{Tempe}
    \state{AZ}
    \country{USA}
}
\email{yinanli2@asu.edu}

\author{Hasti Seifi}
\orcid{0000-0001-6437-0463}
\affiliation{
    \institution{Arizona State University}
    \city{Tempe}
    \state{AZ}
    \country{USA}
}
\email{hasti.seifi@asu.edu}

\begin{abstract}
Mid-air ultrasound haptic technology can enhance user interaction and immersion in extended reality (XR) applications through contactless touch feedback. Yet, existing design tools for mid-air haptics primarily support creating tactile sensations (i.e., tactons) which cannot change at runtime.
These tactons lack expressiveness in interactive scenarios where a continuous closed-loop response to user movement or environmental states is desirable. This paper introduces AdapTics, a toolkit featuring a graphical interface for rapid prototyping of \textit{adaptive tactons}---dynamic sensations that can adjust at runtime based on user interactions, environmental changes, or other inputs. A software library and a Unity package accompany the graphical interface to enable integration of adaptive tactons in existing applications.
We present the design space offered by AdapTics for creating adaptive mid-air ultrasound tactons and show the design tool can improve Creativity Support Index ratings for Exploration and Expressiveness in a user study with 12 XR and haptic designers. %
\end{abstract}



\begin{teaserfigure}
    \centering
    \begin{subfigure}[t]{4.44792cm} %
        \includegraphics[page=1,
        height=4.6cm,
        width=\linewidth, trim = 0.5cm 8.75cm 24.5cm 0.60cm, clip]{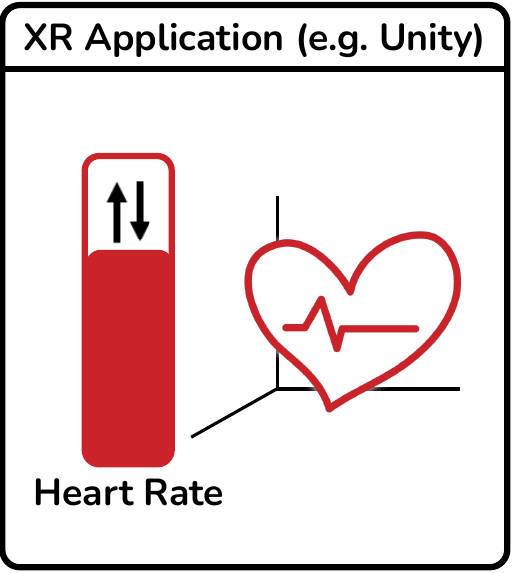}
        \captionsetup{skip=4pt}
        \caption{A virtual heart.}
    \end{subfigure}
    \hspace{1.8mm} %
    \begin{subfigure}[t]{7.116672cm} %
        \includegraphics[page=1,
        height=4.6cm,
        width=\linewidth, trim = 9.5cm 8.75cm 9.5cm 0.60cm, clip]{figures/Teaser_Figure.t.pdf}
        \captionsetup{skip=4pt}
        \caption{Designing an adaptive tacton (e.g., the teal pattern) using ``Heart Rate'' as an external parameter.}
    \end{subfigure}
    \hspace{1.8mm}
    \begin{subfigure}[t]{4.44792cm} %
        \includegraphics[page=1,
        height=4.6cm,
        width=\linewidth, trim = 24.5cm 8.75cm 0.5cm 0.60cm, clip]{figures/Teaser_Figure.t.pdf}
        \captionsetup{skip=4pt}
        \caption{Feeling the adaptive tacton change with Heart Rate.}
    \end{subfigure}
    \caption{AdapTics enables rapid prototyping and refinement of interactive tactile experiences. Starting with an existing virtual environment (a), designers create adaptive tactons with the AdapTics Designer by using environment parameters to control tacton properties (e.g., size, modulation, speed, and sequence) (b), then integrate the adaptive tacton into the virtual scene via the AdapTics Engine (c).}
    \Description[
    Three panels depict the AdapTics workflow: (a) "XR application (e.g., Unity)": shows a heart icon and a "Heart Rate" meter, (b) "AdapTics Designer": illustrates a spatial tactile icon in the shape of a "bow-tie" (center) with adaptations depicted as icons for size, modulation, speed, jumps (right) based on the "Heart Rate" meter (left), and (c)  "XR App + AdapTics Engine" shows the same meter as the first scene with 3 different levels highlighted, showing hands experiencing different sensations depending on the "Heart Rate" meter.
    ]{
    A three-panel sequence illustrating the AdapTics workflow.
    (a) "XR Application (e.g., Unity)": A 3D environment displays a heart icon next to a vertical bar meter labeled "Heart Rate".
    (b) "AdapTics Designer": The "Heart Rate" meter is shown on the left, with an arrow labeled "f(x)" pointing to icons on the right. These icons represent changes in size, modulation, speed, and a "jump" within a sequence of keyframes. In the center of the abstracted Designer interface, there is a set of keyframes laid out in the shape of a "bow-tie" with a cursor creating the "path" for the tacton.
    (c) "XR App + AdapTics Engine": Shows the same meter as the first scene with 3 different levels highlighted. Three arrows from the "Heart Rate" meter points to three variants of haptic feedback with varying size and intensity.
    }
    \label{fig:teaser}
\end{teaserfigure}

\maketitle

\section{Introduction}
Mid-air ultrasound haptic technology offers contactless haptic feedback for a growing list of real-world and immersive applications. With the fast update rate of the technology (e.g., at 20~kHz), one can modulate the position and intensity of ultrasonic focal points to create a wide range of spatiotemporal touch sensations (e.g., shapes, movement, and rhythms) on the user's body~\cite{RakkolainenIsmo2021ASoM}. Users can feel mid-air ultrasound sensations while stationary or moving without needing gloves, wearables, or any physical contact.
Designers are exploring the technology's potential in various applications, including touchless public or virtual displays~\cite{Corenthy2018, maunsbach2022whole}, car dashboard controls~\cite{Harrington2018}, medical training simulations~\cite{Hung2013, Hung2014}, and immersive experiences in virtual reality (VR) environments~\cite{makino2016haptoclone,seinfeld2022evoking} and games~\cite{martinez2018touchless, Villa2022}.

Despite the growing interest in mid-air ultrasound technology, creating interactive or dynamic haptic experiences remains a significant challenge.
While software APIs offer flexibility for designing interactive haptic feedback, programming can hinder rapid prototyping and creative exploration, essential for haptic design~\cite{schneider2017haptician, maclean2017multisensory}, and limit designers without technical backgrounds~\cite{seifi2020novice}.
Graphical haptic design tools provide a more accessible and efficient alternative, but current tools limit designers to creating tactile icons or tactons~\cite{brewster2004tactons} that are \textit{fixed} or \textit{non-adaptive}~\cite{Seifi2023, schneider2016macaron}.
Once designed, such tactons produce haptic output that is predefined and unalterable.
Fixed tactons are adequate for event-based feedback like alerts but have limited use in interactive scenarios that require real-time adjustments to haptic feedback in response to user interaction or application state.
Designers can sequence multiple tactons to create more complex patterns~\cite{Seifi2023}, but this approach does not support continuous adjustments and is cumbersome for complex interactions.
This gap in the design process highlights the need for tools that enable the creation of \textit{adaptive tactons}, which are tactile sensations capable of dynamically changing in real-time in response to environmental inputs.
In the rest of the paper, we use the terms fixed and non-adaptive interchangeably to refer to the predefined, unalterable nature of non-adaptive tactons.

Unlike fixed tactons, adaptive tactons offer designers the flexibility to adjust the haptic output at runtime in response to one or more inputs.
For example, as a user interacts with a virtual button or knob, a designer can dynamically alter the size of a tacton element to illustrate the degree of activation with \textit{value mapping}.
Designers may also use \textit{conditional triggers} to transition between touch sensations based on an input trigger. In the button example, a different segment of the tacton can be rendered when the button activation reaches 100\%.
In contrast to fixed tactons, which are typically brief and discrete, adaptive tactons lend themselves to being continuously played for the full duration of an interaction.
These capabilities of adaptive tactons enable designers to create dynamic and interactive feedback.
For instance, the designer may create an adaptive heartbeat tacton to replicate the player character's heart rate in a VR game (Figure~\ref{fig:teaser})
or in navigational aids for the visually impaired, adaptive tactons could adjust to indicate changes in terrain or proximity to obstacles.

This paper introduces \textit{AdapTics}, a new open-source haptic design tool and rendering engine to rapidly prototype and efficiently render adaptive mid-air haptic tactons.
The \textit{AdapTics Designer} allows designers to create an ultrasound tacton, adjust the tacton parameters directly, or link them to external parameters using formulas (\textit{value mapping}) or conditional jumps (\textit{conditional triggers}). The designers can then test these adaptive tactons on the ultrasound haptic device using an integrated 3D simulation environment and iterate on their design.
Given the rapid update rate of mid-air ultrasound technology, supporting adaptive tacton playback is challenging since, unlike fixed tactons, adaptive tactons cannot be precomputed and must be evaluated on demand.
Addressing this challenge, the design tool is accompanied by the \textit{AdapTics Engine}, a native application and software library that can evaluate and render adaptive tactons in real-time, an order of magnitude faster than the update rate of typical ultrasound haptic devices. The software library and a Unity package facilitate integrating the adaptive tactons into external applications (e.g., a VR project).

To build \textit{AdapTics}, we iteratively defined a design space for adaptations in mid-air ultrasound tactons and evaluated the tool and design space in a user study with 12 designers.
Inspired by the design practices in adaptive audio~\cite{FMOD, Wwise}, the design space of adaptive ultrasound tactons offers five dimensions that describe the (1) granularity level of adaptations, (2) timing manipulations to speed and temporal sequence of tacton, (3) spatial adaptations to the tacton's size, position, and rotation, (4) feel changes, and (5) the type of transformation or mapping between the external and tacton parameters.
After building this design space in AdapTics, we conducted a user study with 12 XR and haptic designers to evaluate the tool's effectiveness and gather feedback.
Participants designed tactons using the AdapTics design tool as well as a second version of the tool that did not support adaptations and tested the sensations on an ultrasound haptic device.
We used the Creativity Support Index (CSI) psychometric survey~\cite{cherry2014quantifying} to quantitatively measure the efficacy of the adaptive design features in AdapTics.
Our results showed significant improvements in Exploration, Expressiveness, and the overall creativity support score when the adaptive features were present in the tool.
The main contributions of this work are:

\begin{itemize}
\item AdapTics toolkit: an open-source haptic design tool, rendering engine, and Unity package for creating and integrating adaptive mid-air ultrasound haptic experiences.
\item A design space for adaptive mid-air ultrasound tactons to guide and inspire haptic designers.
\item Results of a user study demonstrating the utility of AdapTics for creative design.
\end{itemize}

\section{Related work}
We summarize past research on mid-air haptic technology and applications, then outline the progress in haptic design tools. Finally, we present design practices in adaptive audio and provide evidence on the need for adaptive tactons in haptics. %

\subsection{Mid-Air Ultrasound Haptic Technology and Applications} %

Mid-air ultrasound technology provides spatial, contactless touch experiences by controlling an array of ultrasonic transducers to focus at a point (i.e., a focal point) on the user's skin~\cite{Takayuki2009, hoshi2010noncontact}.
The frequency of the focal point is above 16~kHz, typically 40~kHz, which is not directly perceptible by the human skin~\cite{RakkolainenIsmo2021ASoM}.
Thus, haptic designers must use a modulation technique to stimulate human touch receptors.
Two common modulation techniques include applying a low-frequency envelope (<1000~Hz) to the ultrasound carrier in amplitude modulation or AM~\cite{Takayuki2009, Long2014} or moving the focal point rapidly along a path in spatiotemporal modulation or STM~\cite{frier2018using, frier2019sampling}.
Ultrasound technology can also create multiple focal points, but this approach reduces the tactile sensation's strength~\cite{Alexander2011, Carter2013} and thus is rarely used by haptic designers.
AdapTics supports both AM and STM as well as the combinations of these techniques for a single focal point, providing the highest level of control over ultrasound parameters in a graphical haptic design tool to date.

Mid-air ultrasound haptic technology offers a large set of parameters for designing tactons. Like mechanical vibrations, ultrasound haptics provides temporal parameters for design such as AM frequency, duration, intensity, and rhythm~\cite{obrist2013talking,dalsgaard2022user}.
In addition, designers can move the ultrasound focal point to render various haptic shapes and paths in 3D.
In spatiotemporal modulation, the focal point's path and movement speed (i.e., drawing frequency) can be modulated over time.
The size, rotation, and position of the ultrasound shapes are also important design parameters in mid-air haptics. RecHap showcases the large design space of the technology with a library of 500 hand-designed mid-air ultrasound tactons, augmented 20 times to a total of 10,000 sensations~\cite{theivendran2023rechap}. The combination of tacton parameters can influence the perceived intensity of the ultrasound~\cite{frier2019sampling}, user performance in recognizing different haptic patterns~\cite{Hajas2020}, as well as user emotions and associations for the ultrasound signals in an application~\cite{obrist2015emotions,dalsgaard2022user}.
AdapTics facilitates the exploration of this large design space with a graphical interface, allowing designers to quickly experiment and iterate their designs.

While mid-air ultrasound haptic feedback is typically applied to the hand using a single stationary device~\cite{Corenthy2018, maunsbach2022whole, hwang2017airpiano, martinez2018touchless}, others have extended the use cases of the technology by using multiple devices in tandem~\cite{Suzuki2019}, attaching commercial or custom ultrasound arrays on a fixed stand or a VR headset to render haptics on the face~\cite{gil2018whiskers} or mouth~\cite{Shen2022}, or mounting the device to a robotic arm to enable room-scale haptic feedback~\cite{Villa2022}.
Ultraleap has commercialized mid-air ultrasound technology, providing a hardware platform for research and application development.
The AdapTics toolkit has built-in support for Ultraleap devices, such as the STRATOS Explore~\cite{STRATOSExplore}, and can support other custom ultrasound configurations with a flexible and generalizable API for hardware communication.

\subsection{Design Tools for Haptics}
Recent studies have exposed the complexities of haptic signal design and the need for software tool support in haptics~\cite{schneider2017haptician,maclean2017multisensory,seifi2020novice,Seifi2023}. These studies highlighted the importance of rapid prototyping and iterative refinement for designers~\cite{schneider2017haptician} and the difficulty of programming haptic devices for novices~\cite{seifi2020novice}. Programming and debugging ultrasound patterns are especially challenging due to the large design space of ultrasound parameters and the low intensity of contactless haptic sensations with the technology~\cite{Seifi2023}.

Graphical design tools facilitate haptic design by supporting rapid exploration and prototyping of tactons. Several GUI tools exist for authoring tactons for different technologies, such as vibration actuators~\cite{schneider2016macaron,lee2009vibrotactile,schneider2015tactile,paneels2013tactiped,pezent2020_syntacts}, force-feedback knobs~\cite{swindells2006tool}, and pneumatic jackets~\cite{delazio2018force}. These tools highlight the importance of direct manipulation for creating and refining tactons~\cite{maclean2017multisensory} and promote a shared set of features such as a library of examples, a timeline, easy access to haptic playback, and a visual preview of tactons.
For mid-air ultrasound haptics, Ultraleap provides a graphical interface called Sensation Editor for accessing a library of about 20 example tactons with predefined control parameters (e.g., circle radius) in their development kit. DOLPHIN provides a graphical interface for selecting a geometric shape from a set of primitives such as circle or arc and helps researchers systematically sample their spatiotemporal parameters (e.g., drawing frequency) for psychophysics studies~\cite{mulot2021dolphin}. Feellustrator is a recent GUI design tool for prototyping mid-air ultrasound tactons, developed based on interviews with mid-air haptic designers~\cite{Seifi2023}.

Current haptic design tools primarily produce fixed, non-adaptive tactons, limiting expressiveness in interactive scenarios. For instance, in VR applications, tactile feedback must reflect the specific timing and manner of user actions, such as when pressing a virtual button or moving through a virtual environment, to provide continuous reinforcement. Fixed tactons, which cannot dynamically change with user interaction, can cause confusion or diminish immersion.
In other cases, the designer might be able to create multiple fixed tacton versions to accommodate variations (e.g., different hand sizes), but this can quickly become unmanageable with more interactive elements.
While other tools like Feellustrator focus on non-adaptive tacton design, AdapTics aims to enable both the design and real-time rendering of adaptive tactons, catering to the evolving demand for expressivity and interactivity in XR haptic design.

\subsection{Adaptive Stimuli Design in Haptics and Other Modalities}
Haptic researchers have previously reported the need for adapting tactons at runtime. Past studies showed that users can miss vibrotactile notifications depending on their daily physical activities (e.g., walking vs. biking)~\cite{Blum2015}, emotional state~\cite{Fortin2019}, or the location of the notification on the user's body~\cite{karuei2011detecting}.
These studies highlighted the need for adjusting tacton parameters (e.g., intensity) based on the user's state before delivering the notification~\cite{Fortin2019}.
Others showed the importance of responsive haptic signals during user interaction. Specifically, Sabnis et al. found that a tight coupling between the tactile feedback and user movement can change the user perception and association of a haptic effect from a distal (i.e., remote external event) to a proximal one (i.e., the result of the user's action)~\cite{sabnis2023tactile}.
To enable proximal haptic feedback, Sabnis et al. presented Haptic Servos, a vibrotactile device and an algorithm for mapping user input (e.g., moving a slider) to the firing rate of a vibration pulse~\cite{sabnis2023haptic}.
Degraen et al. presented Weirding Haptics, a vibrotactile tool for creating a vibration tacton for a virtual object with voice and adapting the vibration amplitude and frequency at runtime based on the speed and position of the user's hand~\cite{degraen2021weirding}. To simplify design for novices, Weirding Haptics provided three types of adaptations: looping a tacton, changing the tacton's frequency and amplitude linearly with user speed, or fixing the frequency and amplitude to position landmarks.
This system is the closest to our work yet focuses on mechanical vibrations and a subset of possible adaptations.
In contrast, AdapTics aims to support adaptations across the entire design space and tackles challenges unique to mid-air ultrasound haptics, including the computational complexity arising from more design parameters, and the need to meet hardware update rates (in kHz) for ultrasound rendering, while ensuring responsive adaptations with low latency.

Adaptive audio design practices can inform the field of adaptive haptic design, serving as a precedent for constructing dynamic sensory experiences.
Developed to enhance player immersion and reduce auditory repetition in gaming~\cite{Bossalini2021}, adaptive music employs audio middleware like FMOD~\cite{FMOD} and Wwise~\cite{Wwise} to adjust the audio track at runtime.
These tools enable parameter-based adjustments, such as modifying the strength of a low-pass filter in relation to a game character's underwater depth.
Another common feature is controlling the music composition with horizontal re-sequencing~\cite{hazzard2019adaptive}, i.e., using looping and jumping to create nonlinear audio sequences.
While adaptive audio utilizes layering, or vertical re-orchestration~\cite{hazzard2019adaptive}, this aspect does not translate well to ultrasound haptics, as dynamically adding and removing focal points can drastically impact the sensation's strength.
Both parameter-based adjustments and horizontal re-sequencing informed the design space of adaptive mid-air tactons for AdapTics.

In summary, AdapTics addresses the limitations of current haptic design tools in ultrasound haptics, which are confined to fixed, non-adaptive tactons. Recognizing the need for more versatile haptic feedback, particularly in mid-air ultrasound haptics where the vast design space and contactless interaction heighten demand for such mechanisms, AdapTics enables the creation of advanced, contextual, and dynamically changing feedback, similar to the concept of adaptive audio in video games. This includes adaptations like controlling tacton speed to signify urgency or progress, accommodating different hand sizes, and ensuring seamless transitions in sync with visual feedback.

Building on the above haptic and audio design literature, we present a design space and a software toolkit for creating adaptive mid-air tactons in the following sections.

\section{Design Space of Adaptive Mid-Air Ultrasound Tactons in AdapTics} \label{sec:designspace}

\begin{figure*}[b]
    \centering
    \includegraphics[width=\linewidth]{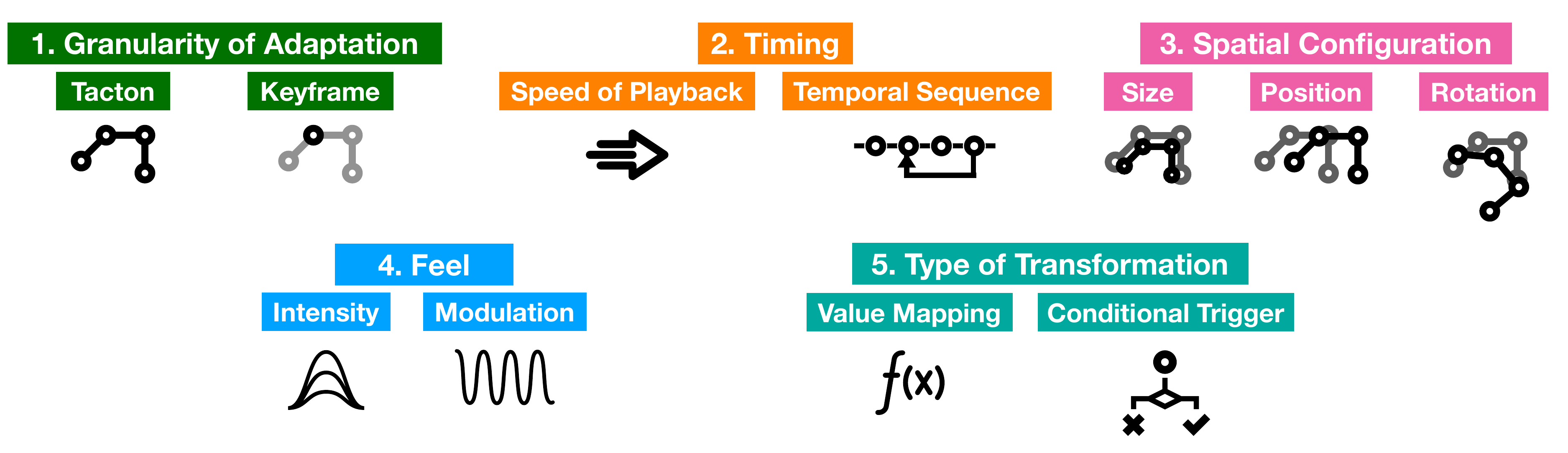}
    \caption{The design space of adaptive mid-air ultrasound tactons in AdapTics with five dimensions and their possible values. An adaptive tacton can have multiple values on each dimension such as adaptations to both size and rotation under spatial configuration (D3).}
    \Description[Each of the 5 dimensions of the design space are laid in groups out with a colored header. An identically colored sub-header and icon are used to describe the values for each dimension.]{Each of the 5 dimensions of the design space are laid in groups out with a colored header. An identically colored sub-header and icon are used to describe the values for each dimension. Under Granularity of Adaptation (D1) a set of 4 circles (keyframes) are connected with lines for the "Tacton" (1a), and for "Keyframe" (1b) only a single circle is highlighted. Under Timing (D2) "Speed of Playback" (2a) is depicted as an arrow with a "motion blur" trail. "Temporal Sequence" (2b) is shown as a connected line of circles (keyframes), with an additional arrow looping backwards. Under Spatial Configuration (D3), the same icon used for "Tacton" (1a) is shown scaled down, translated right, and rotated clockwise, each from a "ghost" copy of the original version, to describe "Size" (3a), "Position" (3b) and "Rotation" (3c) respectively. Feel (D4) shows a half wave with different amplitudes for "Intensity" (4a), and a sine wave for "Modulation" (4b). Type of Transformations (D5) uses an "f(x)" icon for "Value Mapping" (5a). and a decision tree with failure or success for "Conditional Trigger" (5b).}
    \label{fig:designspace}
\end{figure*}

AdapTics provides a design space with five dimensions for adaptive ultrasound tactons~(Figure~\ref{fig:designspace}). The first four dimensions describe the tacton parameter(s) that are adapted at runtime according to: (D1) \textit{Adaptation Granularity}, (D2) \textit{Timing}, (D3) \textit{Spatial Configuration}, and (D4) \textit{Feel}.
The fifth dimension, \textit{Type of Transformation}, describes the mapping between the external variables (e.g., speed of a virtual object) and tacton parameters.
A single adaptive tacton can use one or more values in each dimension.
We iteratively defined this design space by reviewing example haptic interactions in the literature, exploring design practices in adaptive audio, and brainstorming with our team. This design space can support designers in exploring and comparing various adaptive tactons for a target use case before committing to a design. %

\textbf{1. Adaptation Granularity:}
This dimension describes where the runtime adaptations are applied in the tacton with two alternatives: (1a) \textit{tacton} where the runtime adaptations affect the rendering parameters of the ultrasound pattern as a whole, and (1b) \textit{keyframe} where the adaptations affect the rendering parameters of individual keyframes (i.e., a subset of the tacton). Here, a tacton is composed of one or more keyframes, thus along this dimension, the designer can modify the tacton at two levels of granularity depending on the use case.
For example, when designing a tacton for an ``ocean wave'', a designer might scale the entire tacton to match the size of a user's palm~\cite{salagean2022virtual} or opt for keyframe adaptation to intensify just the crest of this wave.

\textbf{2. Timing:}
This dimension describes runtime manipulations to the (2a) \textit{speed of playback} and (2b) \textit{temporal sequence} of a tacton. With the \textit{speed of playback}, a tacton can be rendered faster or slower according to external parameters. For example, when turning a virtual knob, the detents can play faster or slower depending on the user speed~\cite{degraen2021weirding,Villa2022}. The \textit{temporal sequence} of rendering the tacton can also change by specifying a jump to a timestamp in the tacton if a condition is met. For example, the designer can loop to the start of a tacton while the user is in an interaction state (e.g., being in the proximity of a button) or jump to another part of the tacton to reflect transitions between interaction states (e.g., transition to pressing the button)~\cite{maunsbach2022whole}.

This dimension is inspired by current practices in designing adaptive audio ~\cite{sweet2015interactivemusic,collins2017pac}. Specifically, changing the speed of playback is similar to changing the tempo of a soundtrack, and conditional jumps in the tacton's temporal sequence are similar to the transitions between different soundtracks (i.e., horizontal re-sequencing) in adaptive audio.

\textbf{3. Spatial Configuration:}
Along this dimension, designers can adapt the (3a) \textit{size}, (3b) \textit{position}, and (3c) \textit{rotation} of a tacton at runtime. For example, a tacton can guide the user's hand toward a target (e.g., an object or pose) by changing its size, position, and orientation according to the hand's distance and direction to the target ~\cite{Mulot2023}. As another example, the designer can create a tacton representing a ``falling leaf''~\cite{Villa2022} and adapt its size, position, and orientation on the user's hand according to the graphical representation in a VR scene.

\textbf{4. Feel:}
Besides the timing and spatial configuration of the tacton, the designer can adapt the qualitative sensation of the tacton with (4a) \textit{intensity}, and (4b) \textit{ultrasound modulation} of the tacton. The \textit{ultrasound modulation} refers to adapting the frequency of the amplitude modulation~\cite{hoshi2010noncontact} or the drawing frequency~\cite{frier2018using,frier2019sampling} of the tacton. The haptics literature provides examples of this adaptation for vibrotactile tactons. For example, the intensity of a tacton can increase when user's attentiveness or tactile sensitivity is lower due to movement or emotions~\cite{Blum2015,Fortin2019}. As another example, changes to the AM frequency of a tacton can convey that the user is squeezing or pressing a deformable virtual object~\cite{degraen2021weirding}.

\textbf{5. Type of Transformation:}
This dimension describes how an external parameter is mapped to the tacton parameter(s) with two options: (5a) \textit{value mapping}, (5b) \textit{conditional trigger}. In \textit{value mapping}, the value of the external parameter is mapped to the value of a tacton parameter (e.g., AM frequency, speed of playback) directly or with a mathematical formula. In contrast, in \textit{conditional trigger}, the value of the external parameter is used to determine whether to jump to a different part in the tacton or not. Here, the external parameter can represent the state of the user (e.g., hand size~\cite{salagean2022virtual} or physiological state~\cite{Fortin2019}), application (e.g., speed of a virtual object), or interaction (e.g., how far a button is pressed~\cite{maunsbach2022whole}).

In this design space, each of the \textit{Adaptation Granularity} (D1) and \textit{Type of Transformation} (D5) must have at least one value for the tacton to be adaptive. From the other three dimensions (D2, D3, D4), at least one dimension must have a value for the tacton to be adaptive. In the next section, we describe how the AdapTics toolkit supports this design space.

\section{AdapTics Toolkit}

We developed AdapTics as an online open-source toolkit for designing adaptive mid-air ultrasound tactons.
The design of AdapTics was informed by the literature and existing design tools in haptics and adaptive audio as well as by the authors' experience in mid-air ultrasound design. We reference relevant sources that informed our design decisions inline as we present the toolkit below.
The toolkit and its source code can be accessed at this address: \url{https://github.com/AdaptiveHaptics}.

\subsection{Overview and Design Considerations}
The AdapTics design toolkit has two core components: the Designer and the Engine.
This section provides an overview of their functionalities, unique features, and architectural choices.

\begin{figure*}[h]
    \centering
    \begin{subfigure}[c]{0.75\linewidth}
        \includegraphics[width=\linewidth,page=1]{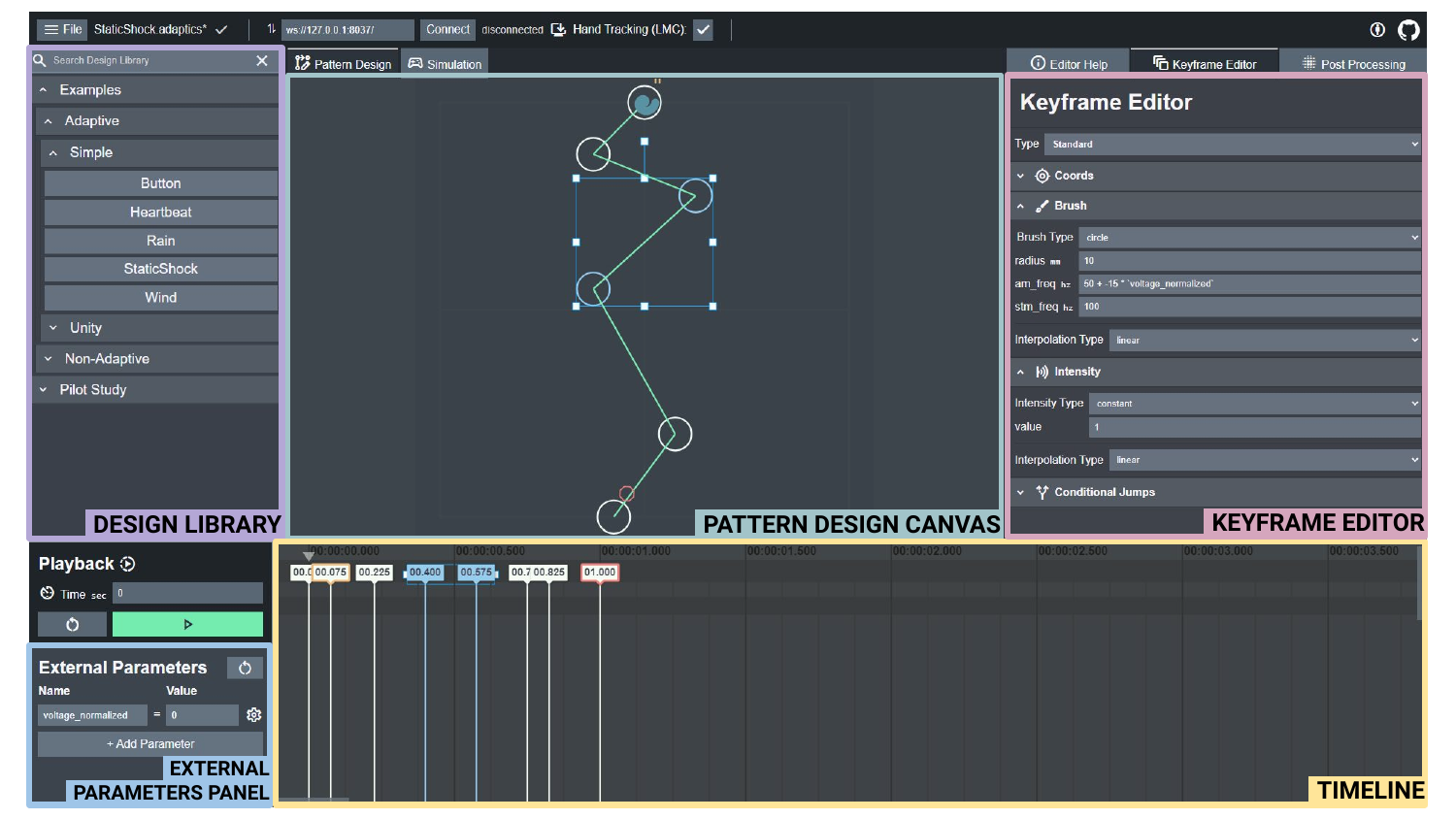}
    \end{subfigure}
    \begin{subfigure}[c]{0.23\linewidth}
        \includegraphics[width=\linewidth,page=2,clip,trim=17.53cm 4.95cm 0.5cm 0.8cm]{figures/AdapTics_Diagrams.t.pdf}
    \end{subfigure}
    \caption{The AdapTics Designer's web interface showing the Design Library, Pattern Design Canvas, Keyframe Editor Tab, External Parameters Panel, Timeline Pane, and the Post Processing Tab.}
    \Description{(a) A labeled screenshot of the AdapTics Designer interface. The left-center contains a "Design Library" with a directory structure and search box. The main pane in the center is the "Pattern Design Canvas" where the tacton's path is created. On the right-center, the "Keyframe Editor" tab displays properties of selected keyframes in categories like "Coords", "Brush", "Intensity", and "Conditional Jumps". The bottom-left corner showcases the "External Parameters Panel", listing parameter names and their values. The remainder of the bottom area presents the "Timeline" with the keyframe sequence of the current tacton.
    (b) Reveals another tab from the right-center pane: "Post Processing", organized in a two-column layout without tiers.
    The interface's top bar (unlabeled) includes a "File" menu, the design's name, and configuration options.
    }
    \label{fig:designer}
\end{figure*}

The \textit{AdapTics Designer} is an online graphical tool for creating and modifying adaptive tactons (Figure~\ref{fig:designer}).
The Designer draws inspiration from the design requirements and concepts established by Feellustrator for non-adaptive tacton design~\cite{Seifi2023} with elements such as the pattern design canvas, timeline, and design library.
The tool introduces new functionality and features for creating adaptive tactons such as support for parametric design and an integrated 3D environment to test and debug adaptations, drawing from and extending best practices in adaptive audio~\cite{FMOD, Wwise}.
The Designer also provides new features for improving non-adaptive tacton design, such as supporting AM, STM, and combinations of their techniques and offering different transition types between keyframes to enable a wider range of tactons, better catering to the needs of haptic designers~\cite{RakkolainenIsmo2021ASoM, Seifi2023}.
The Designer was built as a web based application to better support cross platform use across operating systems and device form factors (e.g., desktop, laptops, tablets). This decision was also informed by the ease of access and large-scale deployment and analytics afforded by web-based haptic tools~\cite{schneider2016macaron,seifi2019haptipedia}.
We developed the Designer with vanilla JavaScript, leveraging the Konva.js~\footnote{\url{https://github.com/konvajs/konva}} library for its 2D graphical design panes and Three.js~\footnote{\url{https://github.com/mrdoob/three.js}} for the 3D simulation environment.
To render haptic feedback on the ultrasound hardware and receive hand-tracking data, the Designer relies on a network connection to the AdapTics Engine, but is capable of independent operation when ultrasound hardware or the Engine are unavailable, a feature requested by mid-air ultrasound designers~\cite{Seifi2023}.

The \textit{AdapTics Engine} is primarily responsible for rendering tactons on the ultrasound haptic hardware.
Unlike non-adaptive tactons, where the location and amplitude of the ultrasound focal point can be precomputed, adaptive tactons require real-time evaluation.
Additionally, to keep the feedback delay of adaptations minimal, the ``batch'' or buffer size of focal points must be kept small.
Thus, the Engine must ensure consistent performance---free from disruptions like garbage collection and minimizing heap allocations---to meet the fast update rate of the ultrasound hardware and keep adaptation latency low. %
To accomplish this, the Engine is predominantly written in Rust~\cite{matsakis2014rust}, a systems programming language designed for memory and thread safety, with C and C++ bindings to interface with hardware device SDKs.
The Engine's architecture can support any body-tracking and ultrasound hardware configuration. This hardware independence is important due to the diversity and ongoing developments in the ultrasound haptics and motion tracking~\cite{RakkolainenIsmo2021ASoM,caserman2019survey,buckingham2021hand}.
Therefore, for robustness and flexibility, the Engine calculates tacton time exclusively relative to the ultrasound device's time, allowing for devices with variable update rates or operating in environments without a reliable external time source.
This approach supports \textit{timing adaptations (D2)} in our design space, allowing for conditional jumps and adaptive changes to playback speed, even accommodating negative speeds.
The Engine includes a native command line application that facilitates bi-directional communication between the Designer and ultrasound hardware via WebSockets.
For rendering AdapTics tactons in other applications, the Engine provides a software library with a C-compatible API.

\subsection{Non-Adaptive Tacton Design Features in AdapTics}

In AdapTics, tactons consist of a sequence of \textit{keyframes}. In the AdapTics Designer, keyframes can be created and reordered through the Pattern Design Canvas and Timeline Pane (Figure~\ref{fig:designer}).

Various attributes of a keyframe or the entire tacton can be changed within the Keyframe Editor and Post Processing tabs, respectively.
The Brush attribute describes what the haptic device renders as the path is traversed.
The user can select the brush type (e.g., circle or line~\cite{frier2018using}) and adjust brush-specific parameters, such as size, rotation, and the frequencies for both AM and STM as well as adjust the relative strength with Intensity.
Users can select either ``linear'' or ``step'' interpolation between keyframes for the Coordinates, Brush, and Intensity properties. Linear interpolation ensures a smooth transition, while step interpolation creates a distinct change between adjacent keyframes. These transition types were informed by our own design experience and various use cases in mid-air ultrasound haptics such as presenting a sequence of disconnected points on the user's palm~\cite{dalsgaard2022user,pittera2019m}.

In the Post Processing Tab, the user can adjust the playback speed, scale the intensity, or apply geometric transformations (rotate, scale, translate) to the entire tacton.
The Post Processing Tab's utility especially emerges when creating adaptive tactons.

These functionalities replicate and improve upon existing tools for non-adaptive ultrasound tacton design~\cite{Seifi2023}. %

\subsection{Adaptive Tacton Design through External Parameters}
External parameters are what enable designers to create adaptive tactons.
These parameters allow parts of the tacton to change in real-time, responding to outside events or states.
Parameters have a name, defined when the tacton is being designed, and a value updated by an external application (e.g., Unity) at runtime.
In the Designer, the user can create external parameters in the External Parameters Panel and manually change their value to test adaptations within the design environment.

In AdapTics, the external parameters can directly change the value of one or more tacton attributes (i.e., \textit{value mapping} in Section~\ref{sec:designspace}).
Nearly all numeric fields in AdapTics permit using parameters and mathematical formulas.
We integrated formulas into AdapTics to encapsulate the full scope of adaptive design within the tacton itself. For instance, instead of directly setting a parameter like \texttt{rumble\_frequency} to 50 from the external application, a high-level parameter such as \texttt{health} can be mapped to the appropriate AM frequency within the tacton. %
Supporting formulas was requested by mid-air ultrasound designers but not included in prior design tools~\cite{Seifi2023}.
Formulas can support any number of parameters and basic arithmetic operations.
Only the time and coordinates of a keyframe must be constant numbers. Instead, the designer can parameterize the playback speed or geometric transformations of the path in the Post Processing Tab.

Another use for external parameters is in conditional jumps (i.e., \textit{conditional trigger} in Section~\ref{sec:designspace}). %
Conditional jumps are an optional property of keyframes and allow the flow of the tacton to be controlled, making it possible to branch to another section of the tacton based on some condition.
Currently, AdapTics Designer allows for specifying one or more conditions that compare an external parameter against a fixed number, jumping to a timestamp if the condition is met.

In summary, external parameters, value mapping, and conditional jumps allow the user to create tactons that are almost entirely parametric.
The design and implementation of external parameters in AdapTics, for example, the ability to manually control parameter values during testing and the representation of jump targets as 'flags' in the timeline, draws from established practices in adaptive audio design tools~\cite{FMOD, Wwise}.

\subsection{3D Simulation Environment}
Adaptive tactons' dynamic nature necessitates the modulation of external parameters for effective testing.
Manually adjusting more than one parameter in the External Parameters Panel can be cumbersome, and using external applications like Unity for iterative refinement is time-consuming, as it involves frequent switching between interfaces and costly scene launches.

To streamline the design process for haptics~\cite{schneider2017haptician,maclean2017multisensory}, we introduce a 3D Simulation Environment integrated directly into the AdapTics Designer. This environment serves as a stand-in application, enabling the modulation of tacton parameters without requiring the full standalone counterpart.
It features a base environment with a representation of the ultrasound device, a visualization of the ultrasound focal point trail, and a 3D hand model for tracking purposes.
With this feature, the user can craft custom 3D scenes for testing adaptive tactons directly within the AdapTics Designer, though some coding is required.

\begin{figure*}[t]
    \centering
    \includegraphics[width=.9\linewidth]{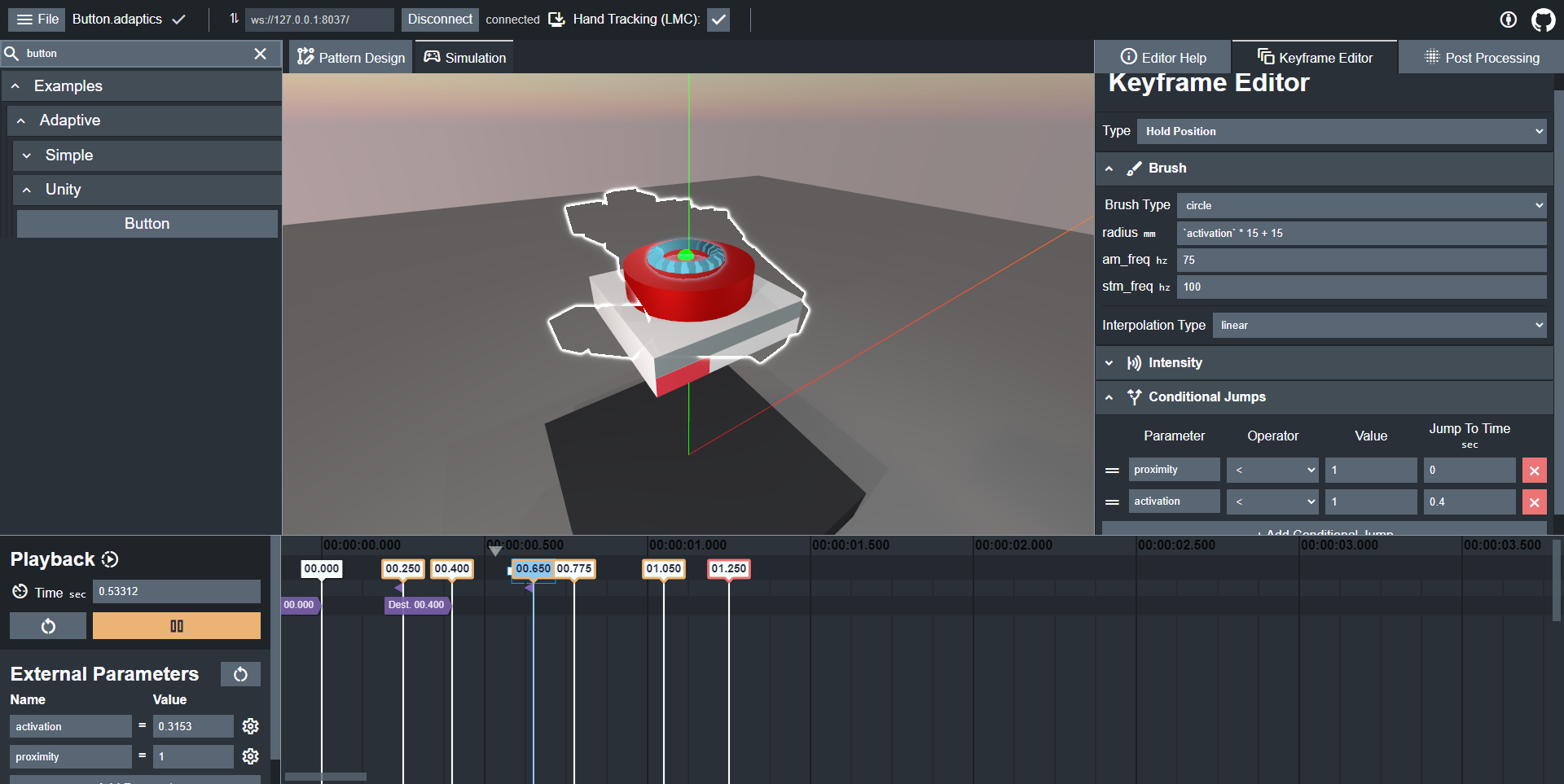}
    \caption{AdapTics Designer's integrated 3D simulation environment showcases the testing of an adaptive tacton for a button. The Button scene uses hand tracking to update the values of the external parameters in real-time, visible in the External Parameters Panel. Also, the Keyframe Editor showcases using the parameters both within formulas and in conditional jumps.}
    \Description{AdapTics Designer interface with the center pane showing the 3D Simulation tab. In the 3D space, a matte box represents the ultrasound device. Above the ultrasound device is a floating button. A low poly outline of a hand can be seen pressing this button with the palm. A 3D tube visualization of the haptic sensation is shown on the hand. Currently it is a circle, with a gradient representing the amplitude modulation in this screenshot. In the Keyframe Editor, a parameter can be seen in a formula on the brush>radius property (activation * 15 + 15). Parameters can also be seen in the conditional jumps section of the Keyframe Editor: if `proximity` `<` `1` jump to `0` OR if `activation` `<` `1` jump to `0.4` }
    \label{fig:3dpanel}
\end{figure*}

To support this, we provide both an extendable base environment and several interactive example scenes with their scripts. Each interactive scene introduces 3D elements that respond to the user's hand movements, updating the tacton's external parameters accordingly. For example, the Button scene updates two parameters, \textit{proximity} and \textit{activation} based on interactions with a virtual button (Figure~\ref{fig:3dpanel}).

\subsection{Application Integration, Hardware Compatibility, and Performance of the AdapTics Engine}
\begin{figure}[h]
    \centering
    \includegraphics[width=0.95\linewidth]{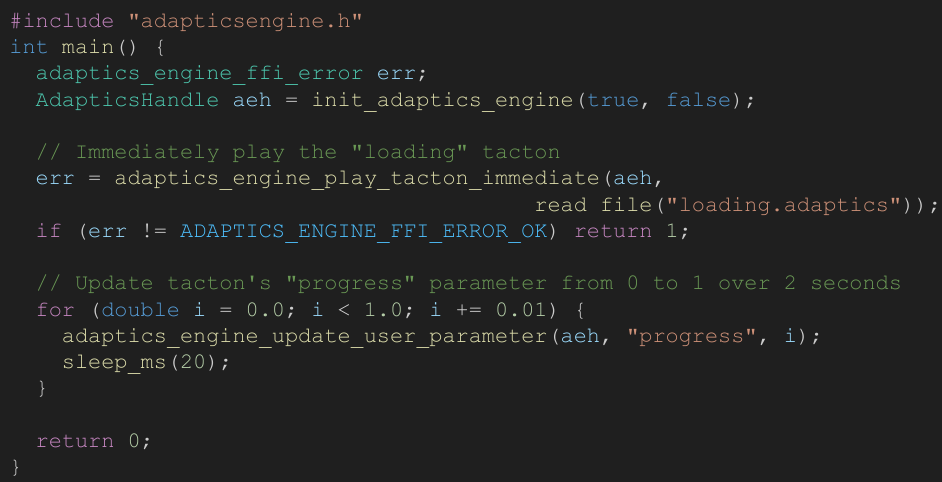}
    \caption{Example usage of the AdapTics Engine API in C showcasing functions for (1) loading and playing the tacton, and (2) updating the tacton with the ``progress'' external parameter.} %
    \Description{C code including the adapticsengine.h, initializing the engine with `init_adaptics_engine`, playing a tacton from a file 'loading.adaptics' with `adaptics_engine_play_tacton_immediate`, and updating a parameter in a for loop (i from 0 to 100) with `adaptics_engine_update_user_parameter('progress', i)` and a call to `sleep_ms(20)` to emulate a longer loading time. Please see figure's embedded text for full code.}
    \label{fig:capicode}
\end{figure}

With a few lines of code, the AdapTics Engine can be integrated into various applications using its C-compatible API.
The Engine provides control over tacton playback, adjustment of external parameters, and even supports hot reloading of tactons during playback.
To ensure ease of integration, the API includes functions for common use cases, such as loading and immediately playing a tacton and updating the tacton parameters, demonstrated in Figure~\ref{fig:capicode}.
Also, the API allows applications to manipulate a geometric transformation matrix to position, scale, or rotate the tacton on a user's skin via body tracking or to have the tacton follow a virtual object (e.g., a bee) in the environment.

\paragraph{Unity Package} To facilitate the integration of adaptive tactons in VR, the AdapTics Engine Unity Package serves as a wrapper for the Engine, with high-level features tailored for Unity applications.
The package includes a prefab with the AdapTics Engine Controller script, acting as the ultrasound coordinate origin.
This prefab contains two child GameObjects: a reference ultrasound device placed relative to the origin and a visualization of the ultrasound focal point's trail.
Notably, the AdapTics Unity Package allows the tacton to follow the position of any GameObject, facilitating the use of any Unity-compatible body tracking system.

\paragraph{Hardware Support} The AdapTics Engine currently has implementations for communicating with Ultraleap ultrasound haptic devices or a mock device when an ultrasound device is unavailable.
The Engine's device-agnostic architecture facilitates adding support for other haptic hardware.
Specifically, the architecture keeps track of conditional jumps and adaptive playback speed changes without relying on a fixed hardware time step or external clock. The Engine can operate at any, even variable, device refresh rates.
\begin{figure}[h]
    \centering
    \includegraphics[width=0.9\linewidth]{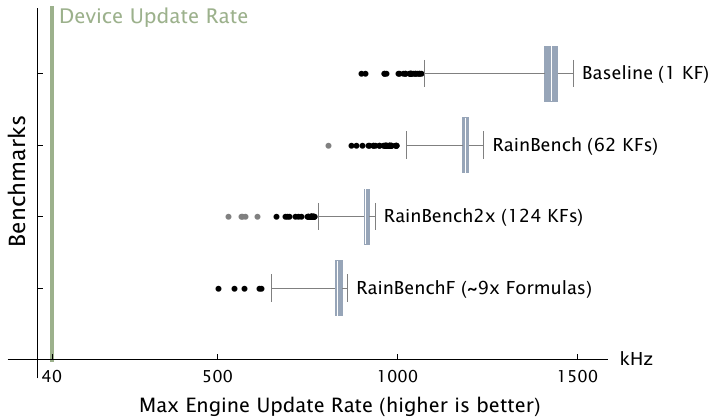}
    \caption{
    Results of a synthetic benchmark profiling the maximum update rate achievable by the AdapTics Engine when evaluating adaptive tactons.
    Starting from the beginning of each tacton, 1000 batches of 40 samples were evaluated as quickly as possible, giving a theoretical maximum update rate for the focal point.
    The tactons had between 1 (\textit{Baseline}) and 124 keyframes (\textit{RainBench2x}).
    The \textit{RainBenchF} tacton has the same number of keyframes as \textit{RainBench}, but has $\sim$9~times more formula computations.
    }
    \Description{
    Box plot of the benchmark results for each tacton in kHz. Device update rate at 40 kHz is marked for reference. Median values are approximately 1428, 1187, 911, and 831 kHz for Baseline (1 KF), RainBench (62 KFs), RainBench2x (124 KFs), and RainBenchF (∼9x Formulas) respectively. The interquartile ranges and standard deviations are fairly tight, with a range of approximately 30-90 kHz. There is a large range of outliers, however, with the minimum recorded values (including outliers) for each tacton being: 900, 809, 531, 503 kHz, in the same order as before.
    }
    \label{fig:bench}
\end{figure}

\paragraph{Performance}
We ran exploratory benchmarks with complex and large adaptive tactons to gauge AdapTics' tacton evaluation performance margin compared to the update rate of an ultrasound haptic device.
Besides ensuring smooth playback, faster tacton evaluation can also reduce buffer size, minimizing the latency between parameter adjustments and changes in tactile feedback.
We expected that performance is primarily impacted by the total number of keyframes and the number and size of formulas in a tacton, based on code profiling and analysis.
Our benchmark focused on the rate of tacton evaluation, omitting any potential overhead from communicating with the ultrasound hardware, as we aimed to understand the Engine's performance independent of specific hardware implementations.
We ran the benchmark on a workstation with an AMD Ryzen~9 3900X CPU and DDR4-3200 memory, using Windows 10. A Unity application was active during the benchmark, reflecting the AdapTics Engine's typical use-case scenario.

As illustrated in Figure~\ref{fig:bench}, the AdapTics Engine maintained update rates far beyond the rate for current ultrasound hardware. Even with complex tactons, such as \textit{RainBenchF} with 62 keyframes and $\sim$100 formula operations per evaluation, the Engine managed to sustain playback at a rate of 500~kHz or higher, at least 10 times faster than the update rate of current ultrasound hardware (e.g., Ultraleap STRATOS Explore).
Practically, this data suggests the Engine can handle high frequency and low latency playback of even large and complex adaptive tactons.

\section{User Evaluation}
We conducted a user study to evaluate the efficacy of AdapTics as a creativity support tool and compare adaptive and non-adaptive tacton design for haptic and XR applications. The study was approved by the university's ethics review board.  %

\subsection{Participants and Recruiting}
We recruited 12 participants with backgrounds in haptics or VR through snowball sampling. To be eligible, the participants must have completed at least one project with VR or haptic technology. This criterion helped ensure the participants could comprehend and engage with the design tasks and reflect on the utility of adaptive features for their work. Each participant received a \$25 Amazon gift card as compensation.

\subsection{Apparatus and 3D Scenes}
We created two versions of the AdapTics Designer tool to assess how adaptive features impact creative tacton design. Version A supported both non-adaptive and adaptive tacton design, while Version B was limited to non-adaptive tacton design. Specifically, we omitted the External Parameters Panel, the conditional jump feature, and the 3D Simulation Environment to create Version B. Half of the participants started the design tasks with the non-adaptive version (B), while the other half started with the adaptive version (A). Both versions connected to the Ultraleap STRATOS Explore device with an integrated Leap Motion tracker to render the mid-air tactons in the study.

\begin{figure*}[h]
    \centering
    \begin{subfigure}[b]{0.3\linewidth}
        \includegraphics[width=\linewidth]{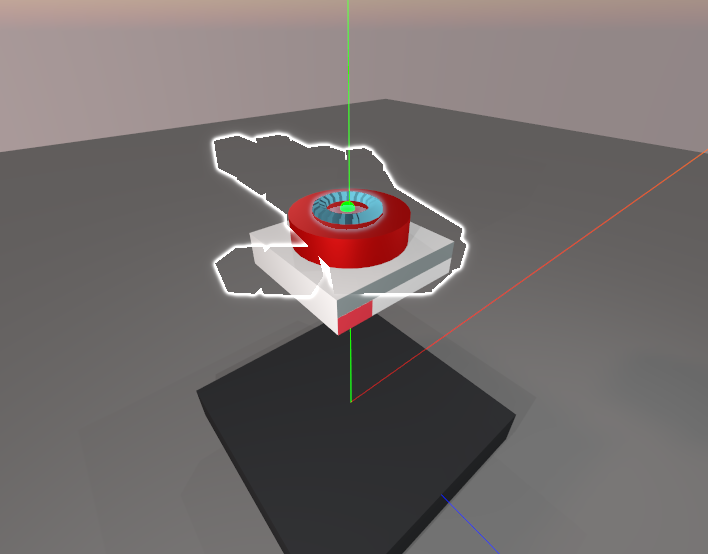}
        \caption{Button scene. The ultrasound tacton changes based on the \textit{proximity} of the user's hand to the button and the \textit{activation} percentage of the button after initial contact. Both parameters vary between 0-1 based on the hand tracking.}
        \Description{A cropped screenshot of the AdapTics Designer's integrated 3D simulation environment is shown. The button is shown floating above the virtual haptic device reference, with an outline hand activating the button about 20 percent. The haptic visualization is shown with a gradient indicating some AM/intensity changes. }
    \end{subfigure}
    \hspace{2mm}
    \begin{subfigure}[b]{0.3\linewidth}
        \includegraphics[width=\linewidth]{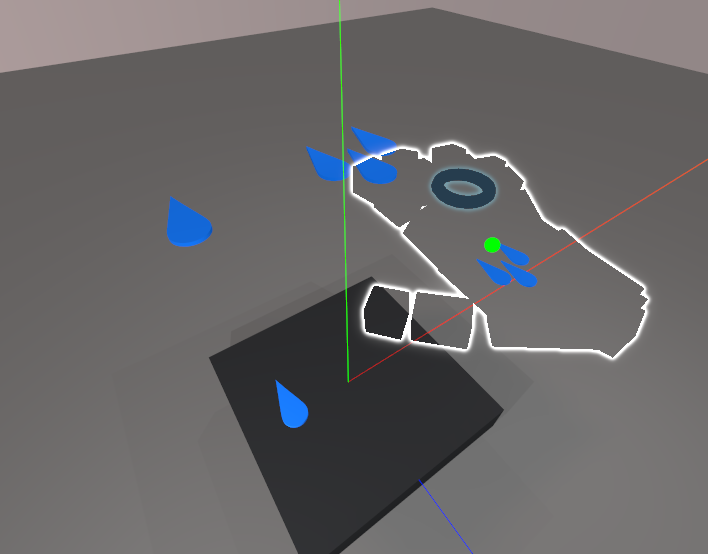}
        \caption{Rain scene. The forward-backward and left-right position of the user's hand controls the \textit{droplet\_strength} and \textit{rainfall\_amount}, respectively. The droplet images provide a visual reference for testing the tacton.}
        \Description{A cropped screenshot of the AdapTics Designer's integrated 3D simulation environment is shown. A set of droplet icons arranged at the corners of a square above the haptic device reference are shown. The bottom left corner shows a single small drop, top left shows a single large drop, top right shows multiple large drops, and bottom right shows multiple small drops. The hand is currently over the multiple small drops icon, but the haptic feedback visualization is dark indicating the intensity of the raindrop is low. }
    \end{subfigure}
    \hspace{2mm}
     \begin{subfigure}[b]{0.3\linewidth}
        \includegraphics[width=\linewidth]{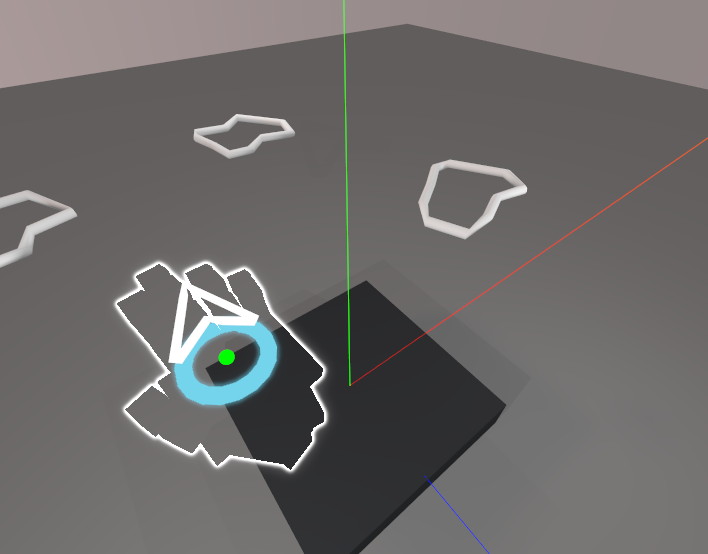}
        \caption{Spaceship scene. The position of the spaceship (white arrow) is controlled with the user's hand. When a flying asteroid hits the spaceship, the value of \textit{taking\_damage} changes from 0 to 1, and the \textit{health} variable decreases.}
        \Description{A cropped screenshot of the AdapTics Designer's integrated 3D simulation environment is shown. The hand is shown above an "arrow" representing the spaceship, with various "asteroids" flying towards it in a 2D plane above the ultrasound device.}
    \end{subfigure}
    \caption{The 3D scenes for the user study. For the Button scene, we designed an adaptive tacton with conditional jumps and value mapping and used the scene as the warm-up for adaptive design tasks. Participants created adaptive tactons for the Rain and Spaceship scenes in the study.}
    \label{fig:3dscenes}
\end{figure*}

We prepared three 3D scenes to allow designers to test the adaptive tactons during the user study, each representing a different type of interaction (Figure~\ref{fig:3dscenes}): (a) \textit{Button} which represents functional interactions with UI widgets, (b) \textit{Rain} showcasing passively feeling virtual sensations, and (c) \textit{Spaceship} depicting interactions within the context of a VR game. The \textit{Button} scene included a single push button with two variables of \textit{proximity} and \textit{activation}.
The \textit{Rain} scene had two variables \textit{droplet\_strength} and \textit{rainfall\_amount} and showed raindrops with different sizes and densities at the four corners of a virtual Ultraleap device. %
Finally, the \textit{Spaceship} scene simulated a simple game where the user could control the position of a spaceship with their palm to avoid flying asteroids, using two variables: \textit{health} and \textit{taking\_damage}.
We selected these interactions and scenes to help participants understand the concept of adaptive tactons and provide them with different design possibilities in the study.

\subsection{Study Procedure}
We conducted the study sessions either remotely via Zoom video calls or in person at our laboratory. To recruit participants with experience in haptic or VR design, we allowed designers who had access to a STRATOS Explore device to participate remotely. Each session took approximately 75-90 minutes.

The session started with background questions, followed by the tool demonstration and warm-up tasks. The background questions asked about the participant's occupation, educational background, and prior experience in VR or haptic design. Next, the experimenter demonstrated the Version B tool (non-adaptive features) to Group 1 and the Version A tool (all features) to Group 2. When participants from Group 1 transitioned to designing adaptive tactons, they received an additional demonstration of Version A.
The participants completed several warm-up tasks such as opening a non-adaptive tacton from the AdapTics library and adjusting its parameters. For the adaptive warm-up task, the participants interacted with the Button scene and explained how the two external variables \textit{proximity} and \textit{activation} modified the tacton at runtime.

Next, the participants created non-adaptive and adaptive tactons for two open-ended design tasks. The participants could feel the ultrasound sensations on their palm throughout the design process, refining each pattern until they achieved the desired tactile feedback. Group 1 first created two non-adaptive tactons with prompts: ``create a pattern that feels like rain'', and ``create a pattern that feels like a heartbeat''. After being introduced to the adaptive features (Version A), the participants interacted with the Rain and Spaceship scenes to see how the variables were updated, %
then designed a rain tacton for the Rain scene and a heartbeat tacton for the Spaceship scene. %
They were free to utilize the scene variables as they wanted to make their tactons adaptive. Conversely, Group 2 began with designing the adaptive tactons and later transitioned to designing their non-adaptive counterparts. The participants shared their screens and described their thoughts.
In the closing interview, we asked about what worked or did not work in the tool, any other adaptive features the participants needed, and the pros and cons and use cases of adaptive vs. non-adaptive tactons. %

\paragraph{Data collection} We collected participant ratings for the tools using the Creativity Support Index~(CSI)~\cite{cherry2014quantifying} and recorded all the study sessions. CSI is a psychometric survey to assess how well a tool can support creative work along six dimensions: Collaboration, Exploration, Expressiveness, Immersion, Enjoyment, and Results Worth Effort. The survey has 12 agreement statements, each rated on a scale of 1 (Highly Disagree) to 10 (Highly Agree), and 15 paired-factor comparisons where the user is presented with all pairs of the six factors and selects the most important factor in completing the creative task. The participants completed the 12 agreement ratings after interacting with each version of the tool (A and B). Following the guidelines for administering CSI, the agreement statements were displayed in a random order, were not grouped by the factors, and did not include the factor names. Upon completing all four design tasks, the participants completed the paired-factor comparison test.
We also video-recorded user interactions with the design tools and their responses to the interview questions and collected all the tacton design files.

\paragraph{Data analysis}
Our data analysis comprised three parts.
First, the experimenter examined tacton designs through video recordings, noting participants' design processes from conceptualization to the creation of loops and parameter adjustments. We analyzed design variations, tacton complexity, and recurring motifs, supported by screenshots of the 48 tactons in AdapTics. These tactons were grouped by design task to identify common patterns and participant preferences.
Subsequently, interview data was transcribed using Otter.ai~\footnote{\url{https://otter.ai/}} and subjected to thematic analysis~\cite{terry2017thematic}. The experimenter and another author separately applied detailed open-coding to the transcripts using MAXQDA~\footnote{\url{https://www.maxqda.com/}}, then discussed the codes and potential themes. Next, the experimenter recoded the transcripts for consolidated analysis and refined the themes with input from the second coder.
Finally, we used one-way repeated measures ANOVAs to analyze the ratings from the CSI questionnaires. This quantitative analysis complemented the qualitative insights, providing a better understanding of the participants' experiences. We present the results from these three analyses below.

\section{Results of User Evaluation}
We first summarize participant backgrounds and the tactons designed in the study, then report the quantitative results of the CSI ratings and the qualitative findings based on the interviews.

\subsection{Summary of Participant Backgrounds}
{
\ifpdf
    \newcommand{\minimidrule}{\specialrule{0.1pt}{0.18ex}{0.26ex}}
\else
    \newcommand{\minimidrule}{\hline} %
\fi
\begin{table*}[t]
    \footnotesize
    \centering
    \begin{tabularx}{\textwidth}{
    >{\centering}m{1.2cm}
    >{\raggedright\arraybackslash}m{2.15cm}
    >{\raggedright\arraybackslash}m{4cm}
    >{\raggedright\arraybackslash}m{4cm}
    >{\raggedright\arraybackslash}m{2cm}
    }
    \toprule
    \textbf{Participant} & \textbf{Education} & \textbf{Haptics Background} & \textbf{XR Background} & \textbf{Other Modalities} \\
    \midrule
    $P1_{XH}$ & PhD in Eng. & Several months in vibrotactile and pneumatic actuators & Several months in VR design with Unity \& Unreal Engine for manufacturing applications & 8 yrs. in audio design \\
    \minimidrule
    $P2_{X}$ & PhD in CS & - & 3 yrs. in medical VR/AR research & Graphic design \\
    \minimidrule
    $P3_{H}$ & Postdoc in CS & 9 yrs. in vibrotactile, force-\allowbreak feedback devices, pneumatic actuators and mid-air ultrasound haptics& - & - \\
    \minimidrule
    $P4_{X}$ & PhD in CS & - & 4 yrs. in locomotion-related VR application design with Oculus SDK and Steam SDK & - \\
    \minimidrule
    $P5_{XH}$ & Bachelor's in Arts & 2 yrs. in vibrotactile design & 2 yrs. in integrating haptic components into VR design with Unity & 2D/3D animation \\
    \minimidrule
    $P6_{XH}$ & Master's in CS & Several months in vibrotactile design & 1 yr. in VR/AR application design & 2D/3D animation \\
    \minimidrule
    $P7_{X}$ & Master's in CS & - & 2 yrs. in VR game development with Unity & - \\
    \minimidrule
    $P8_{XH}$ & Master's in Arts & Several months in vibrotactile gloves and force-feedback devices & Several months in VR design with Unity & 2 yrs. in graphic design \\
    \minimidrule
    $P9_{X}$ & Bachelor's in Eng. & - & 1 yr. in VR design with Unity & 2D/3D animation \\
    \minimidrule
    $P10_{X}$ & Master's in Eng. & - & Several months in VR/AR application design & -  \\
    \minimidrule
    $P11_{H}$ & Master's in Eng. & Several months in vibrotactile for mobile phones & - & 2D/3D animation \\
    \minimidrule
    $P12_{X}$ & PhD in CS & - & Several months in AR/MR design for manufacturing applications & - \\
    \bottomrule
    \addlinespace[1.5mm]
    \end{tabularx}
    \caption{Summary of participant backgrounds. The participant id denotes their design background using a subscript: `H' for haptics, `X' for XR, and `XH' for both XR and haptics. The abbreviations in the table include Eng. for Engineering, CS for Computer Science, and yr for year. }
    \label{tab:participants}
\end{table*}
}

The participants varied in their educational and design backgrounds (Table~\ref{tab:participants}). We had one postdoctoral researcher, four Ph.D. students, five master's students, and two undergraduate researchers, primarily from engineering or computer science (n=10), with others from arts and media (n=2). Participants reported experience in haptics (n=6), VR/AR (n=10), audio design (n=1), graphic design (n=2), and 2D/3D animation design (n=4). %

\subsection{Tacton Design Strategies}
All participants could complete the four non-adaptive and adaptive design tasks. In the non-adaptive tasks, participants employed diverse spatial layouts and ultrasound parameters, using 3 to 83 keyframes (mean: 14).
For the Rain task, most participants (8 out of 12) randomly placed nodes while others placed nodes vertically to simulate the feel of sliding raindrops. For Heartbeat, some overlaid several nodes on the Canvas and varied brush intensity or size to the feel of rhythmic pulses.
The participants used various parameters such as brush type and size, AM frequency, intensity, and speed. In the adaptive tasks, participants followed the same design ideas from the non-adaptive tasks but added adaptive features. Most participants (9 out of 12) used conditional jumps to reduce the number of nodes and eliminate repetitions from their non-adaptive tactons, with resulting tactons ranging between 1 to 42 keyframes (mean: 8). %
Overall, the participants used the conditional jumps for two primary purposes. First, the loops allowed the tacton to continue playing while certain conditions were met (e.g., health > 0). Second, the loops could send the pattern back or forward to different parts depending on the Rain or Spaceship conditions. To make the designs adapt in real-time, participants used the external variables (\textit{health}, \textit{taking\_damage}, \textit{rainfall\_amount}, \textit{droplet\_strength}) directly or with formulas to adjust one or more of the following key attributes: brush size, AM frequency, intensity, and speed. For example, $P1_{XH}$ mapped the \textit{rainfall\_strength} to the ultrasound intensity and the \textit{rainfall\_amount} to the brush size, whereas $P6_{XH}$ mapped the \textit{rainfall\_amount} to the tacton's speed. For adaptive Heartbeat, $P1_{XH}$ mapped the \( 1-\textit{health} \) to the brush intensity, while $P3_{H}$ mapped \( \frac{\textit{taking\_damage}}{\textit{health}} \) to AM frequency.

\subsection{Quantitative Ratings of AdapTics}
\begin{figure*}[t]
    \centering
    \begin{subfigure}[t]{0.45\linewidth}
        \includegraphics[width=\linewidth]{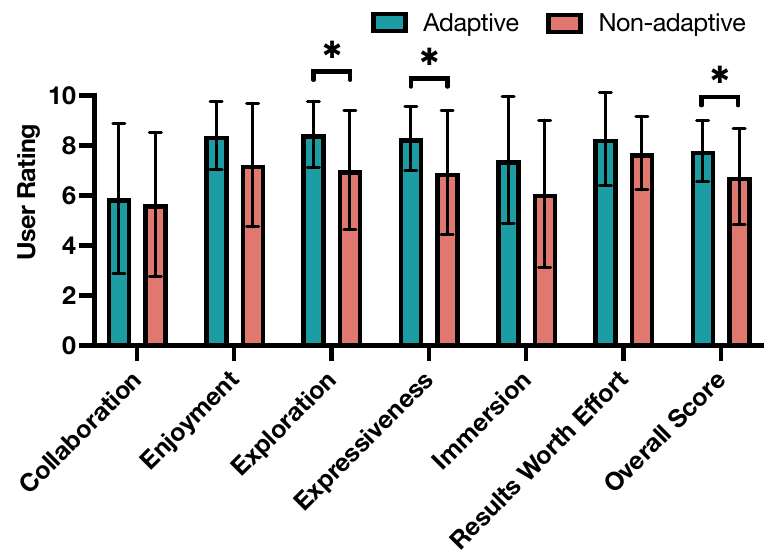}
        \caption{User ratings of the adaptive and non-adaptive tool versions on the six CSI factors and the overall score.}
        \Description{Bar chart comparing the user ratings between the adaptive and non-adaptive tool. User ratings are on a scale from 0-10. The mean scores (generally around 7 to 8.5) for the Adaptive version are higher than the non-adaptive in all categories, but are only significantly higher in Exploration, Expressiveness, and Overall Score, which are called out with stars. Collaboration has a noticeably lower score than the other factors (approximately 6).}
        \label{fig:ratings}
    \end{subfigure}
    \hspace{1mm}
    \begin{subfigure}[t]{0.5\linewidth}
        \includegraphics[width=\linewidth, trim=3.2cm 15.3cm 3cm 3.2cm, clip]{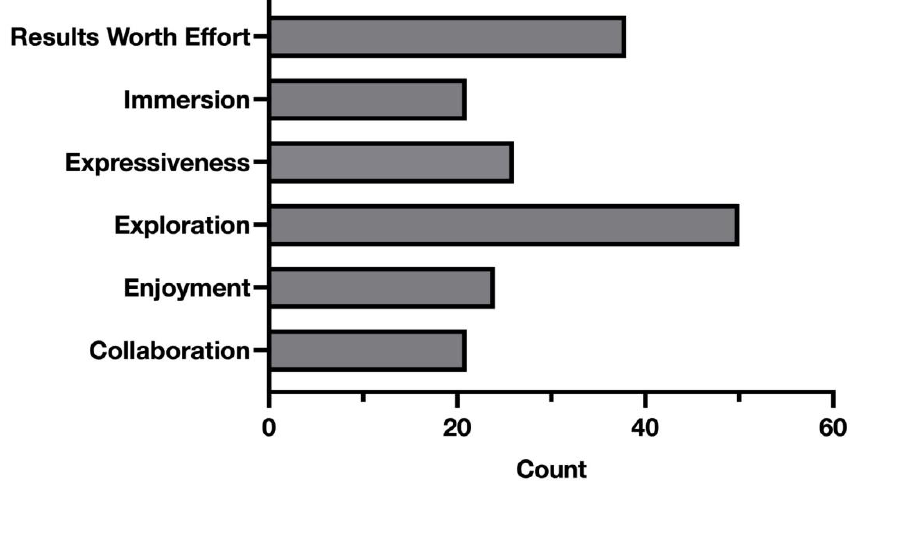}
        \caption{CSI results for the pair-wise comparison task. Factors with higher counts are more important for designers.}
        \Description{6 CSI pair-wise comparison factors, and the counts for each factors, are shown in a bar chart. Most tasks received around 20, but Exploration (approx. 50) and Results Worth Effort (approx. 40) are significantly higher.}
        \label{fig:paired}
    \end{subfigure}
    \caption{Results from the Creativity Support Index (CSI) questionnaire in our user study. }
    \label{fig:plots}
\end{figure*}

Figure~\ref{fig:plots} shows the results of the CSI ratings and paired-factor comparisons. The adaptive tool had higher average ratings than the non-adaptive version on all six factors and the averaged overall score (Figure~\ref{fig:ratings}). In the paired-factor comparisons, participants selected Exploration most frequently as an important factor when designing haptics, followed by Results Worth Effort and Expressiveness (Figure~\ref{fig:paired}).

We used one-way repeated measures ANOVAs to compare the CSI ratings for the \textit{adaptive} and \textit{non-adaptive} tool versions on the six CSI factors and the overall score. Note that the CSI ratings were on a continuous scale from 1-10 with 0.1 increments on the slider, providing interval variables. The assumptions of sphericity and normal distribution held for the data. We present the results at 5\% significance level.

Table~\ref{tab:anova} summarizes the ANOVA results.
The test showed significant differences in the participant ratings for Exploration, Expressiveness, and Overall Score with large effect sizes ($\eta_{p}^{2}\geq.14$). According to the ratings, the adaptive version of the tool supported the Exploration of different design ideas better (mean=8.46, std=1.32) than the non-adaptive version (mean=7.02, std=2.37). Similarly, the adaptive version allowed the participants to be more Expressive (mean=8.29, std=1.28) than non-adaptive (mean=6.92, std=2.47). Finally, the Overall Score was significantly higher for the adaptive tool (mean=7.79, std=1.22) than the non-adaptive version (mean=6.76, std=1.92). Ratings for the other factors did not significantly differ between the tool versions.

\begin{table}[b]
    \centering
    \begin{tabular}{cccc}
         \toprule
         CSI Factor& $F(1,11)$ & \textbf{$P$} & \textbf{$\eta_{p}^{2}$}\\
         \midrule
         Collaboration&.270&.614&.024\\
         Enjoyment&4.159&.066&.274\\
         \textbf{Exploration}&\textbf{7.217}&\textbf{.021}&\textbf{.396}\\
         \textbf{Expressiveness}&\textbf{7.620}&\textbf{.019}&\textbf{.409}\\
         Immersion&4.669&.054&.298\\
         Results Worth Effort&1.312&.276&.107\\
         \textbf{Overall Score}&\textbf{8.201}&\textbf{.015}&\textbf{.427}\\
         \bottomrule
         \addlinespace[1.5mm]
    \end{tabular}
    \caption{Results of repeated measures ANOVA for six CSI factors and the averaged overall score. Test results for Exploration, Expressiveness, and Overall Score showed a significant difference between ratings for the adaptive and non-adaptive versions of AdapTics at $P<.05$ level.} %
    \label{tab:anova}
\end{table}

\subsection{Qualitative Results from Interviews}
Our coding process generated 111 codes, divided into 9 higher-level categories: advantages of adaptive tactons (n=12), applications for adaptive (n=12) and non-adaptive (n=16) tactons, learning cost of the tool (n=20), tool's effectiveness (n=18) and shortcomings (n=5), suggestions for tool enhancement (n=14), long-term design costs for adaptive tactons (n=7), and user study limitations (n=7). We identified 6 themes from these code categories: (1) Tool's effectiveness; (2) Tool's shortcoming and suggestions for enhancement; (3) Applications of non-adaptive tactons; (4) Applications and advantages of adaptive tactons; (5) Learning cost; and (6) Long-term design costs of adaptive tactons. User study limitations are discussed separately in the discussion section.

\paragraph{\textbf{Participants liked the tool's support for rapid prototyping and its adaptive features.}} Several participants described the tool was easy to use for prototyping ultrasound tactons (n=9). $P1_{XH}$ and $P5_{XH}$ liked the intuitive layout of the tool: \emph{``it feels like a professional tool. ($P5_{XH}$)''} $P2_{X}$ valued the instant feedback for parameter tuning and keyframe adjustments, and $P4_{X}$ thought: \emph{``I would enjoy using it in daily life... if I need to design haptic patterns.''}

The adaptation features and 3D Simulation Environment provided new design possibilities for the participants. When we asked about desired adaptive features for haptic design, five participants %
mentioned the tool's comprehensiveness in meeting their needs: \emph{``I think it's surprisingly complete ($P4_{X}$).''} Both $P6_{XH}$ and $P7_{X}$ thought that they could do more with the adaptive version of the tool: \emph{``The conditional statements give me more feel to explore. ($P6_{XH}$)''} $P10_{X}$ wanted to invest more time to further refine their design given the possibilities presented by the adaptive features. $P3_{H}$ thought the adaptive tool allowed for high-level control of the design and wanted to use the tool to assist programming: \emph{``As a programmer, I think it's very nice to have this tool... if I have to do very quick prototyping, I could use this.''} Moreover, $P10_{X}$ commented that there were many alternative ways to utilize the adaptive features for accomplishing more intricate objectives: \emph{``[The adaptive tool] has lots of degrees of freedom to expose a parameter.''} Similarly, $P1_{XH}$ and $P5_{XH}$ referred to the new degrees of freedom offered by the adaptive features. %

\paragraph{\textbf{Participants wanted further support for conditional jumps, 3D simulation, and other haptic technologies.}} %
Four participants suggested improvements to the timeline and conditional jumps. $P5_{XH}$ noted the limitation of using a single timeline: \emph{``It's kind of messy because everything is linked to time.''} They wanted to use multiple timelines for different segments of a tacton to avoid shifting the keyframes and conditional jump flags when modifying a tacton segment. Similarly, $P1_{XH}$ found multiple timelines useful for displaying two sensations simultaneously.
$P9_{X}$ wanted to use a random variable in the conditional jump to enable the keyframes to transition to random timestamps for patterns that require a degree of randomness. $P10_{X}$ suggested adding a unified interface to track all the conditional logic across various keyframes.

Four participants asked for enhanced visualization capabilities in the 3D Simulation Environment (n=4). $P7_{X}$ found it difficult to imagine how the adaptive tacton will play during the interaction in the 3D view: \emph{``When you're trying to design something from pattern design [canvas], it's very difficult to correlate what's going to happen in simulation [3D environment]... When you can get the hang of the correlation from pattern design to simulation, I can see how A would be more useful than B.''} %
$P2_{X}$ and $P4_{X}$ found it difficult to clearly see the movement of the brush in the trail visualization in the 3D view. $P1_{XH}$ suggested using more colors to highlight changes to the tacton parameters such as using color hue for AM frequency.

Finally, three participants suggested extending the tool to other haptic technologies such as vibrotactile feedback: \emph{``this tool is mainly configured for the mid-air haptics and I would really want it to be used for all haptics I think ($P3_{H}$)''}. %

\paragraph{\textbf{Non-adaptive tactons are useful for simple, repeatable, and passive viewing scenarios.}} Eight participants %
described non-adaptive tactons useful for providing simple repeatable feedback to users.  %
For instance, $P2_{X}$, $P6_{XH}$, and $P12_{X}$ mentioned providing notifications about mode-switching or giving warning or confirmation messages in response to user actions. %
Four participants thought fixed tactons were adequate when the user was largely passive, such as simulating the wind from an air conditioner for a stationery user ($P1_{XH}$). %
$P5_{XH}$, $P8_{XH}$, and $P10_{X}$ would use non-adaptive patterns for scripted animations or cutscenes in video games to enrich the narrative. Repeatable, consistent feedback was another benefit of fixed tactons ($P10_{X}$, $P11_{H}$). %
$P11_{H}$ noted that the repeatable feedback could make fixed tactons easier to remember, %
allowing users to recognize different virtual buttons without visual cues. %

\paragraph{\textbf{Adaptive tactons offer new degrees of freedom to reflect real-world interactions.}} When discussing adaptive tactons, several participants drew analogies to real-world physical interactions (n=5). %
For instance, $P1_{XH}$ mentioned creating a more interactive virtual world where haptic feedback could adjust based on the pressure exerted by the user on a virtual object. $P3_{H}$ found adaptive tactons useful for simulating the feel of various surface textures based on user interaction. $P8_{XH}$ and $P10_{X}$ described varying vibration patterns in scenarios such as shooting a gun, driving on different terrains, or feeling the bowstring tension in archery games.
These examples reflected the capacity of adaptive patterns to offer continuously changing feedback to make virtual experiences %
resemble real-world dynamics. $P4_{X}$ captured this sentiment by stating, \emph{``You never touch a button exactly the same way, right? So the adaptive patterns would provide a much more realistic experience there.''}
The participants thought adaptive patterns could provide richer, more dynamic, and more immersive user experiences or convey more information or emotions (e.g., urgency) to intensify gaming experiences (n=9).  %
$P5_{XH}$ summed up this viewpoint by saying: \emph{``The benefit of doing adaptive features gives you a lot more freedom to make the game or the experience feel like it responds to the player, which I think is incredibly important. Otherwise, you might as well just watch a movie and have like a vibrating chair or something.''}

\paragraph{\textbf{The additional degrees of freedom increased the learning cost for designing adaptive tactons.}} %
$P1_{XH}$ and $P7_{X}$ thought the non-adaptive version provided an accessible entry point for users, calling it a \emph{``very good introductory tool ($P7_{X}$)''}. %
Five participants %
said that the adaptive features had a higher initial learning cost. $P5_{XH}$ noted enjoying the adaptive features after mastering how to use them and reflected that adaptive tacton design might require a shift in the designer's mindset:  %
 \emph{``Someone who is learning how to do this, they are often in the mindset of like... I just want to make this one thing and make it work... so for them to learn how to make these adaptive designs, it takes more work.''} $P10_{X}$ pointed out the learning complexities involved in \emph{``Translating what you want from your idea to the working segment... imagining how the graph is like, [how] the keyframes are going to translate into real life.''} $P1_{XH}$ thought seeing example adaptive tactons from other designers could reduce the learning curve for novices: ``\emph{More examples to see how creatively someone could use this... It'll be inspirational for someone who's new.}''. Similarly, others found it useful to see alternative designs for the Rain and Heartbeat tasks.

\paragraph{\textbf{%
The participants had different opinions about the design effort for adaptive tactons over time.}}
Notably, five participants %
highlighted that the burden and cost of the adaptive design are borne by the developers or designers. For instance, $P3_{H}$ thought the tool reduced the effort of programming adaptive tactons but the additional design choices could potentially overwhelm designers, thereby increasing the design cost and complexity. In contrast, $P2_{X}$ anticipated that the design effort spent would decrease as one gains familiarity with the system: \emph{``It's actually I mean, the learning cost is a fixed cost... That cost is huge at the beginning, [but] the more you use a system, the less the average you have to spend.''}

\section{Discussion}
We reflect on the utility of the design space and AdapTics toolkit for prototyping adaptive ultrasound tactons, then outline future steps for extending support for adaptive design in haptics.

\subsection{Utility of AdapTics}
\paragraph{Design Space and Toolkit} The proposed design space and toolkit provide descriptive and generative power for adaptive mid-air ultrasound tactons. The five dimensions of the design space expose the range of adaptation possibilities for ultrasound tactons, helping specify an adaptive design and think about alternatives. The AdapTics toolkit implements this design space, allowing designers to create and test various adaptations across this space.
The web-based graphical interface of the Designer, the real-time performance of the Engine, along with the software API and Unity package support the rapid prototyping and integration that is essential for haptic design~\cite{schneider2017haptician}.

\paragraph{Evaluating AdapTics}
Our study results suggest the utility of AdapTics in exploring and creating expressive tactons.
Following recommended practices for evaluating HCI toolkits~\cite{ledo2018evaluation,remy2020evaluating}, we focused on studying the creativity rather than the usability of AdapTics, involved domain designers, and employed A/B testing to compare the toolkit against a baseline.
The results of CSI ratings and qualitative themes from the interviews match in our study. Specifically, participants noted that the adaptive features provided more degrees of freedom for exploring alternative designs and creating expressive tactons and also rated these two factors (Exploration and Expressiveness) higher in the CSI survey. The average CSI rating for Results Worth Effort did not show a significant effect of the tool version in our study, reflecting the trade-off between expressivity and learning effort in adaptive tacton design.

\paragraph{Use Cases} While the primary focus of this work was to enable prototyping adaptive rather than fixed feedback, the parametric tacton design enabled by AdapTics has additional benefits.
Adaptive tactons can be designed once and used many times, replacing a set or family of non-adaptive tactons with a single parametric tacton.
This approach also allows for haptic feedback that can vary from interaction to interaction, reducing the monotony of repetitive sensations.
As another use case, adaptive tactons enable designers to include all the haptic feedback for one virtual ``object'' or ``interaction'' in a single file, using conditional jumps to control the current sensation.
These capabilities, demonstrated by AdapTics, underscore the utility and potential of adaptive tactons.

\subsection{Limitations}
We developed AdapTics to extend the design power of graphical tools in haptics (i.e., high ceiling)~\cite{ledo2018evaluation,myers2000past} and address the technical challenges associated with prototyping adaptive mid-air ultrasound tactons.
As such, many avenues exist for usability enhancement and further research.
Here we outline AdapTics' primary limitations, informed by user feedback and our observations.

First, AdapTics supports basic formulas with textual editing, but alternatives like graphical curve editors could make manipulation more intuitive and efficient in the future. %
Second, while the AdapTics Designer supports basic collaboration through sharing design files, the web platform could be further leveraged for advanced collaborative tools~\cite{CollaborativeEditing2018, Gao_Gao_Xiong_Lee_2018, Posner1992}, such as simultaneous tacton editing and synchronous feedback mechanisms.
Third, AdapTics provides utilities for debugging through manual control over external parameters, a 3D Simulation Environment, and Unity-based visualizations of focal point trails, but participant feedback has underscored the need for more advanced visualization tools and capabilities.
While participants were typically able to resolve issues through iterative design and testing in the Designer, there is potential for streamlining the identification and prevention of unexpected behavior.
Fourth, %
AdapTics does not currently support vertical re-orchestration, a common feature in adaptive audio, due to perceptual fidelity concerns in ultrasound haptics.
Due the technology's spatial nature, layering of multiple sensations requires the use of multiple focal points, which are generally avoided by designers as they drastically reduce the sensation's overall intensity. Consequently, practical design support for this was out of scope for AdapTics.
Finally, the toolkit is specialized for mid-air ultrasound haptics. Chosen for its large design space, focusing on the unique capabilities and challenges of this technology helped us develop a comprehensive feature set for adaptive design.

The first two limitations could be addressed with further development, while the latter three present research challenges. Next, we discuss how these limitations translate to implications for future research and tool development for adaptive tactons.

\subsection{Implications for Future Work}

\paragraph{Debugging Adaptive Tactons} Unlike fixed tactons, adaptive tactons must incorporate runtime computations or \textit{logic} to enable their adaptability, which brings about the need for debugging.
Future work could explore how to further support this aspect.
For example, tools could reduce the need for debugging by integrating feedforward techniques~\cite{coppers2019fortunettes,muresan2023using} such as offering previews of future tacton states, aiding designers during the incorporation of adaptations like conditional jumps, but more research is needed on when and how to provide such previews to avoid overwhelming the designer.

\paragraph{Vertical Re-orchestration} The concept of adaptive layering, common in adaptive audio~\cite{hazzard2019adaptive}, could become relevant with future improvements to the ultrasound hardware and intensity. %
Addressing this aspect in future tools goes beyond creating multiple timelines as dynamically adding and removing focal points can create unexpected rendering artifacts. Tackling these issues would require dedicated user perception studies, like research by Shen et al.~\cite{shen2023multi}, to understand the impact of adding and removing focal points at runtime and to develop appropriate algorithms for effective vertical re-orchestration. %

\paragraph{Extending to Other Technologies}
In our study, some participants expressed an interest in expanding AdapTics' capabilities to encompass other forms of haptic feedback, such as mechanical vibrations. This feedback suggests not only the perceived utility of AdapTics but also a broader demand for adaptive design tools across different haptic modalities.
Features in AdapTics, such as conditional jumps within the timeline and our support for parametric adaptations through value mapping, are applicable to other haptic technologies.
We provide the complete source code for the AdapTics toolkit to facilitate the reuse of these elements in future haptic tools.
However, creating a general-purpose haptic design tool remains an open challenge due to the distinct design parameters and perceptual characteristics of each haptic technology~\cite{maclean2017multisensory}.
Another promising avenue is integrating adaptive tacton design tools with existing adaptive audio middleware. Such integration would enable the coordinated design and orchestration of both haptic and audio stimuli to enhance immersive experiences in XR environments and games.

\subsection{Reflecting on Methods for Studying Haptic Design Practices}
Our work contributes a snapshot of the practices and opinions of haptic and XR designers about the utility of adaptive design tools and tactons.
Designers' opinions and practices can change over time as they explore the design space of adaptive tactons. For example, in our study, the participants had different opinions on whether the design effort for adaptive tactons would reduce over time. Haptic researchers have outlined the challenges and design activities of novices when programming force-feedback haptic devices through a longitudinal study~\cite{seifi2020novice}. A future direction would be to study how a designer's effort, process, and mindset for non-adaptive vs. adaptive tactons may evolve over time.

To understand these changes, a mixed-methods approach comprising both quantitative metrics like CSI and qualitative thematic analysis proves useful. The CSI ratings objectively evaluated creativity support and user experience with AdapTics, while the thematic analysis offered deeper insights into designers' attitudes, challenges, and perceived benefits of using adaptive tactons. This approach corroborated findings across different data types and also provided a comprehensive view of the practices and challenges in adaptive tacton design. Longitudinal studies employing a similar methodological framework could track changes in both quantitative metrics, such as Results Worth Effort, and qualitative aspects, like evolving tacton creation strategies.

Also, we used the same two design tasks across the study sessions to enable comparison among the tools in the limited study time. Future studies can complement our results by asking designers to use the tool for their own projects. Since recruiting people with design expertise is difficult for a longitudinal study, one could use a participatory design approach for building haptic toolkits through art and design residencies with XR and haptic designers similar to recent initiatives in building graphical design tools for smart textiles and craftwork~\cite{bourgault2023coilcam,devendorf2023adacad,devendorf2023towards}.

\section{Conclusion}
Drawing inspiration from advances in VR and game design, this paper presents AdapTics, an open-source toolkit that facilitates the graphical design of adaptive tactile sensations.
Through its capabilities, AdapTics empowers interaction design researchers and practitioners to introduce the dynamic and expressive qualities inherent in real-world tactile experiences to XR interactions.
With the growing use cases of XR, we anticipate a broader expansion and maturation of the capabilities and applications of adaptive haptic systems.

\begin{acks}
We thank Viresh Bhurke for piloting the integration of AdapTics into various VR scenes in Unity. We also thank the anonymous reviewers, our colleagues, and the study participants for their input on this project.
This work was supported by research grants from VILLUM FONDEN (VIL50296) and the National Science Foundation (\#2339707).

\end{acks}

\bibliographystyle{ACM-Reference-Format}
\bibliography{references}

\end{document}